\documentclass[twocolumn,preprintnumbers,amsmath,amssymb,superscriptaddress]{revtex4-1}
\usepackage{eqnarray,amsmath,lipsum}
\usepackage{graphicx}
\usepackage{dcolumn}
\usepackage{mathtools}
\usepackage{natbib}

\usepackage{color}
\definecolor{red}{rgb}{1,0,0}

\definecolor{blue}{rgb}{0,0,1}

\definecolor{green}{rgb}{0,1,0}

\setcitestyle{square}
\begin{document}
\preprint{APS}
\author {Gabriel Gomes}
\affiliation{S\~ao Paulo State University (UNESP), IGCE, Departamento de F\'isica, Rio Claro, SP, Brazil}
\author {Lucas Squillante}
\affiliation{S\~ao Paulo State University (UNESP), IGCE, Departamento de F\'isica, Rio Claro, SP, Brazil}
\author {A. C. Seridonio}
\affiliation{S\~ao Paulo State University (UNESP), School of Natural Sciences and Engineering, Ilha Solteira, SP, Brazil}
\author {Andreas Ney}
\affiliation{Institute of Semiconductor and Solid State Physics, Johannes Kepler University Linz, Austria}
\author {Roberto E. Lagos}
\affiliation{S\~ao Paulo State University (UNESP), IGCE, Departamento de F\'isica, Rio Claro, SP, Brazil}
\author {Mariano de Souza}
\email{mariano.souza@unesp.br}
\affiliation{S\~ao Paulo State University (UNESP), IGCE, Departamento de F\'isica, Rio Claro, SP, Brazil}
\title{The Magnetic Gr\"uneisen Parameter for Model Systems}
\vspace{0.8cm}
\begin{abstract}
The magneto-caloric effect (MCE), which is the refrigeration based on the variation of the magnetic entropy, is of great interest in both technological applications and fundamental research. The MCE is quantified by the magnetic Gr\"uneisen parameter $\Gamma_{\textmd{mag}}$. We report on an analysis of $\Gamma_{\textmd{mag}}$ for the classical Brillouin-like paramagnet, for a modified Brillouin function taking into account a zero-field splitting originated from the spin-orbit (SO) interaction and for the one-dimensional Ising (1DI) model under longitudinal field. For both Brillouin-like model with SO interaction and the longitudinal 1DI model, for $ T \rightarrow$ 0 and vanishing field a sign change of the MCE is observed, suggestive of a quantum phase transition. SO interaction leads to a narrowing of the critical fluctuations upon approaching the critical point. Our findings emphasize the relevance of $\Gamma_{\textmd{mag}}$ for exploring critical points. Also, we show that the Brillouin model with and without SO interaction can be recovered from the 1DI model in the regime of high-temperatures and vanishing coupling constant $J$.
\end{abstract}
\maketitle
\date{\today}
\section{Introduction}
\begin{figure}
\centering
\includegraphics[angle=0,width=\columnwidth]{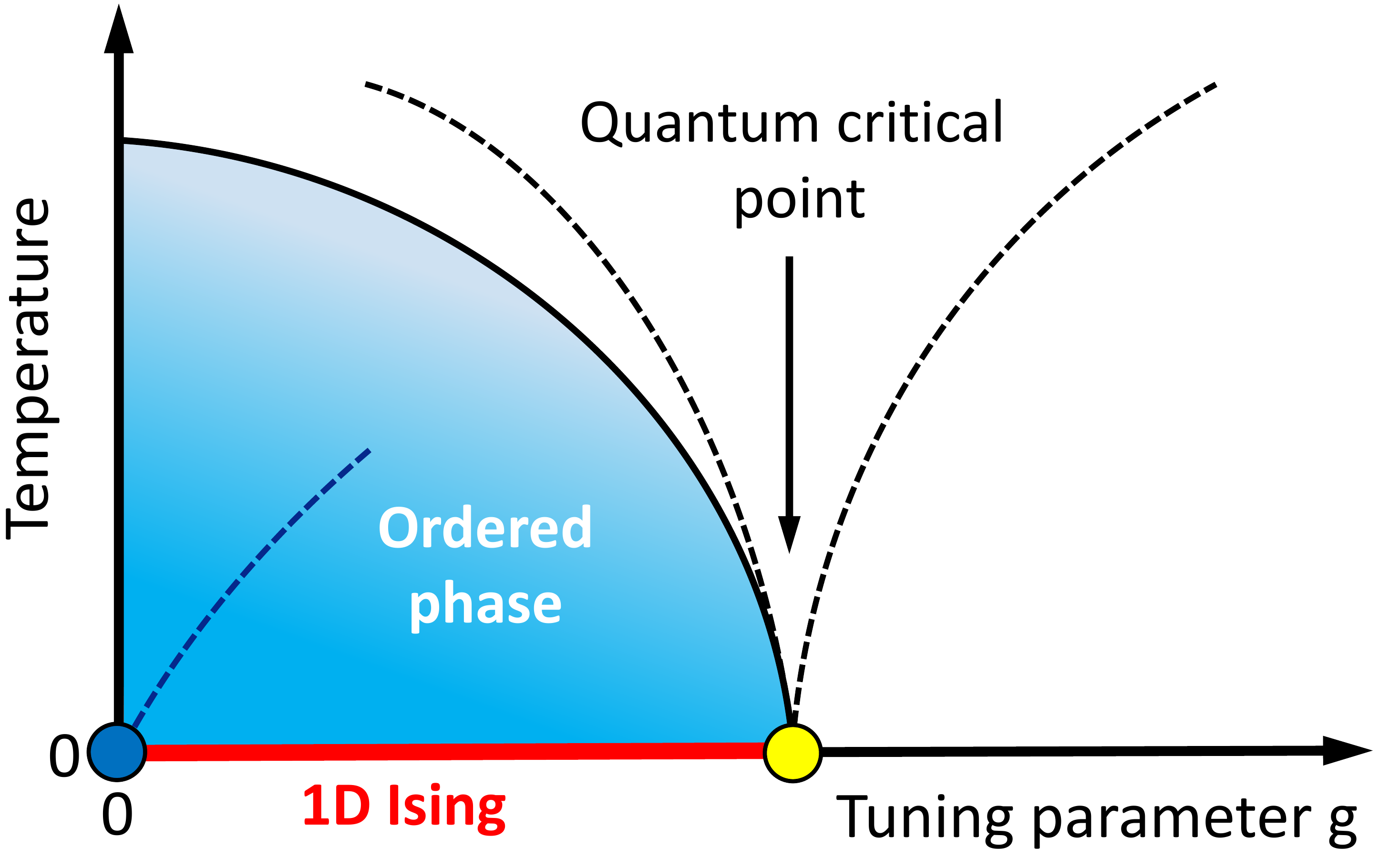}
\caption{\footnotesize Schematic phase diagram of the temperature \emph{versus} tuning parameter $g$,  indicating two quantum critical points. Yellow and blue bullets indicate a magnetic field-tuned and a zero-field QCP, respectively. The red line at $T$ = 0\,K depicts a magnetic field induced QPT for the transverse 1D Ising model, while the dashed lines are associated with crossover temperatures. In the case of a magnetic-field induced QPT the solid line refers to the adiabatic magnetization \cite{honecker}. Figure adapted from Refs.\,\cite{zhu,kinross}.}
\label{Fig-1}
\end{figure}

While \emph{classical} phase transitions are driven by thermal fluctuations \cite{huang}, a genuine \emph{quantum} phase transition (QPT)  \cite{Subir} takes place at $T$ = 0\,K, \emph{i.e.}, thermal fluctuations are absent, and the transition is driven by tuning a control parameter $g$ (see Fig.\,\ref{Fig-1}), namely application of external pressure, magnetic-field or changes in the chemical composition of the system of interest. Intricate manifestations of matter have been observed in the immediate vicinity of a quantum critical point (QCP) (cf.\,Fig.\,\ref{Fig-1}), \emph{i.e.}, the point in which the QPT takes place,  like divergence of the Gr\"uneisen parameter computed by combining ultra-high resolution thermal expansion and specific heat measurements \cite{zhu,kuchler,brando}, collapse of the Fermi surface as detected via Hall-effect measurements \cite{Sylke}, non-Fermi-liquid behavior observed by carefully analyzing the power-law obeyed by the electrical resistivity, specific heat and magnetic susceptibility \cite{Lohne,Flouquet} and breakdown of the Wiedemann-Franz law due to an anisotropic collapse of the Fermi surface \cite{Tanatar}. Hence, the exploration and understanding of the physical properties of interacting quantum entities on the verge of a QCP consists a topic of wide current interest, see e.g.\,\cite{Physics.7.74} and references therein. In this context, heavy-fermion compounds have been served as an appropriate platform to explore such exotic manifestations of matter \cite{Phili}.
Interestingly enough,  new sorts of quantum critical behavior, having strong spin-orbit (hereafter SO) coupling and electron correlations as key ingredients, embodying an antiferromagnetic semimetal Weyl phase \cite{Savary} and excitations of strongly entangled spins called spin-orbitons \cite{Loidl}, have been recently reported in the literature. Furthermore, a QPT in graphene tuned by changing the slope of the Dirac cone has also been reported \cite{Majorana}. In general terms, the fingerprints of a magnetic-field-induced QPT is the divergence of the magnetic susceptibility \cite{Matsumoto} $\chi$($T$,$B$) = $\mu_0$($\partial M$/$\partial B$) (here $M$ refers to the magnetization, $B$ is an external magnetic field, and $\mu_0$ is the vacuum permeability)
for $T \rightarrow$ 0\,K and a sign change of the MCE near QCPs \cite{MGarst}.
Indeed, such fingerprints have been observed experimentally in several materials. Among them YbRh$_2$Si$_2$ \cite{Tokya}, Cs$_2$CuBr$_4$ \cite{Takano} and  CeCoIn$_5$ \cite{Zaum}, just to mention a few examples. Furthermore, zero-field QCPs have been also recently reported in the $f$-based superconductors CeCoIn$_{5}$ \cite{Yoshi}, $\beta$-YbAlB$_4$ \cite{Matsumoto}, and quasi-crystals of the series Au–Al–Yb  \cite{Quasi}. Owing to the experimental difficulties posed by accessing a QCP under pressure and/or under external magnetic-field, zero-field QCPs are of high interest, since quantum criticality is thus accessible simply by means of temperature sweeps. An analogous situation is encountered, for instance, in molecular conductors regarding the finite-$T$ critical-end-point of the Mott metal-to-insulator transition \cite{Mariano,Barto,2015} as well as for gases \cite{EPJ}.
From the theoretical point of view, however, topics still under intensive debate are \emph{i}) the universality class of QPTs \cite{2015,Phili}; \emph{ii}) the temperature range of robustness of quantum fluctuations and the role played by them, for instance, in the mechanism behind high-temperature superconductivity \cite{Physics.7.74}. Several theoretical models have been served as appropriated platforms to address these issues, being the transverse 1DI model, namely an Ising chain under a transverse magnetic field, an appropriate playground to investigate several fundamental aspects \cite{Aeppli,PRB-Ising}. Indeed, for the transverse 1DI model, exactly solved analytically, there is no spontaneous magnetization and a phase transition occurs only at $T$ = 0\,K under finite magnetic field \cite{nolting}, see Fig.\,\ref{Fig-1}. This is merely a direct consequence of the famous Mermin-Wagner theorem \cite{MWtheorem}, which forbids long-range magnetic-ordering at finite temperature in dimensions $d$ $\leq$ 2. Since the 1DI Hamiltonian considers solely nearest-neighbor interactions, computing the eigen-energy of a spin configuration is a relatively easy task \cite{huang}. As a matter of fact, in a broader context, thought the Ising model, at first glance, is a \emph{toy model} to simulate a domain in a ferromagnetic material, it still continues to attract broad interest for instance in the field of quantum information theory \cite{PhysRevA.79.012305} and detection of Majorana edge-states \cite{PhysRevX.4.041028}.
Motivated by the intrinsic quantum critical nature of the 1DI model under transverse field, we explore a possible similar behavior in other exactly analytically solvable models.

This paper is organized as follows: in Section \ref{Brilloui-section} the MCE is calculated for the classical Brillouin-like paramagnet;
in Section \ref{SO-section} the SO interaction is taken into account to calculate the MCE for the Brillouin paramagnet for $S=3/2$, being the results compared with those obtained in Section \ref{Brilloui-section}, the 1DI model under longitudinal field is recalled and the corresponding MCE is presented in Section \ref{Ising-section}.

Before starting the discussions on the MCE for the Brillouin-like paramagnet, it is worth recalling that both 1DI and the two-dimensional Ising (2DI) models provide an appropriate playground to explore critical points both theoretical and experimentally. For zero external field $B$ = 0\,T, the model can be exactly solved and it is known as the famous Onsager solution \cite{Onsager,Slovakei,huang}.  The mathematical solution of both 1DI and 2DI models in the absence of an external magnetic field can be found in classical textbooks, see e.g. \cite{baxter,pathria,nolting}.  A hypothetical sample with volume of 1\,mm$^3$, the SI values of $\mu_B$ = (9.27$\times$10$^{-24}$)\,J.T$^{-1}$, the Boltzmann constant $k_B$ = (1.38$\times$10$^{-23}$)\,m$^2$\,kg\,s$^{-2}$\,K$^{-1}$, and $N$ = (6.022$\times$10$^{23}$)\,atoms were employed in the calculations. For the 1DI model, we have employed a magnetic coupling constant $J$ = 10$^{-23}$\,J = 0.72\,K. Also, it is worth mentioning that the MCE is quantified by the magnetic Gr\"uneisen parameter $\Gamma_{\textmd{mag}}$.
\section{The Brillouin-like Paramagnet}\label{Brilloui-section}
\begin{figure}[!h]
\centering
\includegraphics[angle=0,width=\columnwidth]{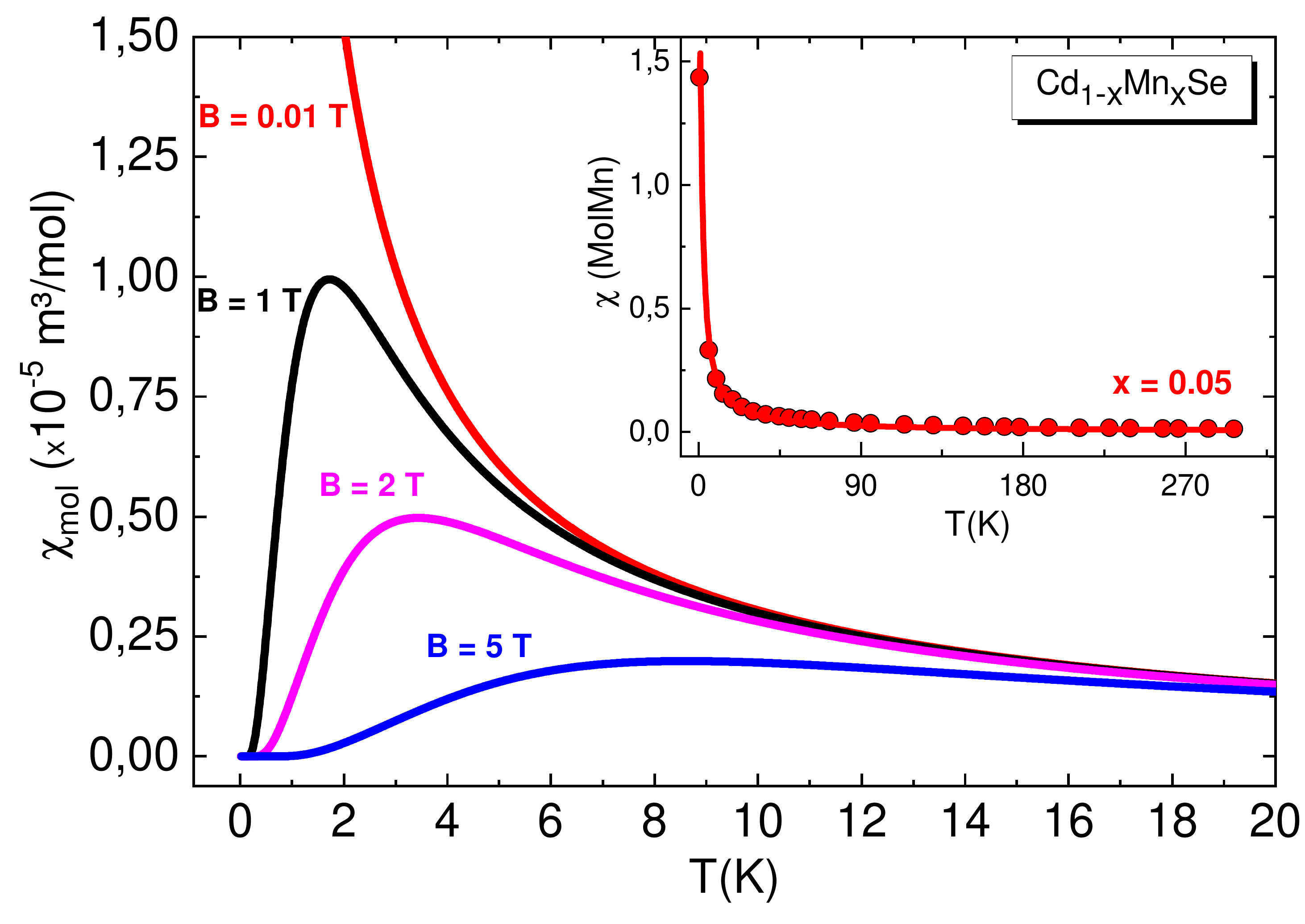}
\caption{\footnotesize Main panel: molar magnetic susceptibility $\chi_{\textmd{mol}} = \mu_0 (\textmd{d}M/\textmd{d}B)$ as a function of $T$ at various magnetic fields for the Brillouin paramagnet employing $J$ = 3/2. Inset: experimental data of the magnetic susceptibility $\chi$ as a function of temperature for the Cd$_{1-x}$Mn$_x$Se paramagnetic system with 5\% concentration of Mn ($x$ = 0.05), data taken from Ref.\,\cite{oseroff}. The red solid line represents a Curie-like fitting, employing the values of $g$ and $J$ of the Cd$_{1-x}$Mn$_x$Se system . The obtained number of spins in the system is $N$ $\sim$ 3$\times$10$^{22}$. Further details are discussed in the main text.}
\label{Fig-2}
\end{figure}

In what follows we discuss the MCE for the Brillouin paramagnet. First, we recall the Brillouin function $B_J$, well-known from textbooks \cite{nolting}:
\begin{equation}
B_J(J,y)  = \frac{2J+1}{2J}\textmd{coth}\left(\frac{2J+1}{2J}y\right)-\frac{1}{2J}\textmd{coth}\left(\frac{y}{2J}\right)
\label{eq:Brillouin-function},
\end{equation}
where $y = g_J\mu_B JB/ k_B T$, $g_J$ is the Land\'e gyromagnetic factor ($g_J$ = 2.274), $\mu _B$ is the Bohr magneton and $J$ the system's spin. The magnetization is readly written as follows:
\begin{equation}
M = N g_B \mu_B J B_J(J,y).
\end{equation}
The magnetic susceptibility is computed by $\chi=(\partial M / \partial B)_{B=0}$.
Figure\,\ref{Fig-2} depicts the Brillouin magnetic susceptibility as a function of temperature under various magnetic fields. Remarkably, at low-$T$ for vanishing magnetic field $\chi$ diverges, as it occurs for a magnetic field-induced QCP. Since $\chi$ and the entropy ($S$) are connected through the free-energy, a divergence of $S$ is also expected at $T$ = 0\,K.
It turns out that for real systems, magnetic moments are always interacting. Such interaction is rather small when compared with the thermal energy $k_B T$, but it becomes relevant at low-$T$ and thus a long-range magnetic ordering takes place \cite{finn1993thermal}. 
The calculation of the MCE for the Brillouin paramagnet is straightforward.
\begin{figure}
\centering
\includegraphics[width=\columnwidth]{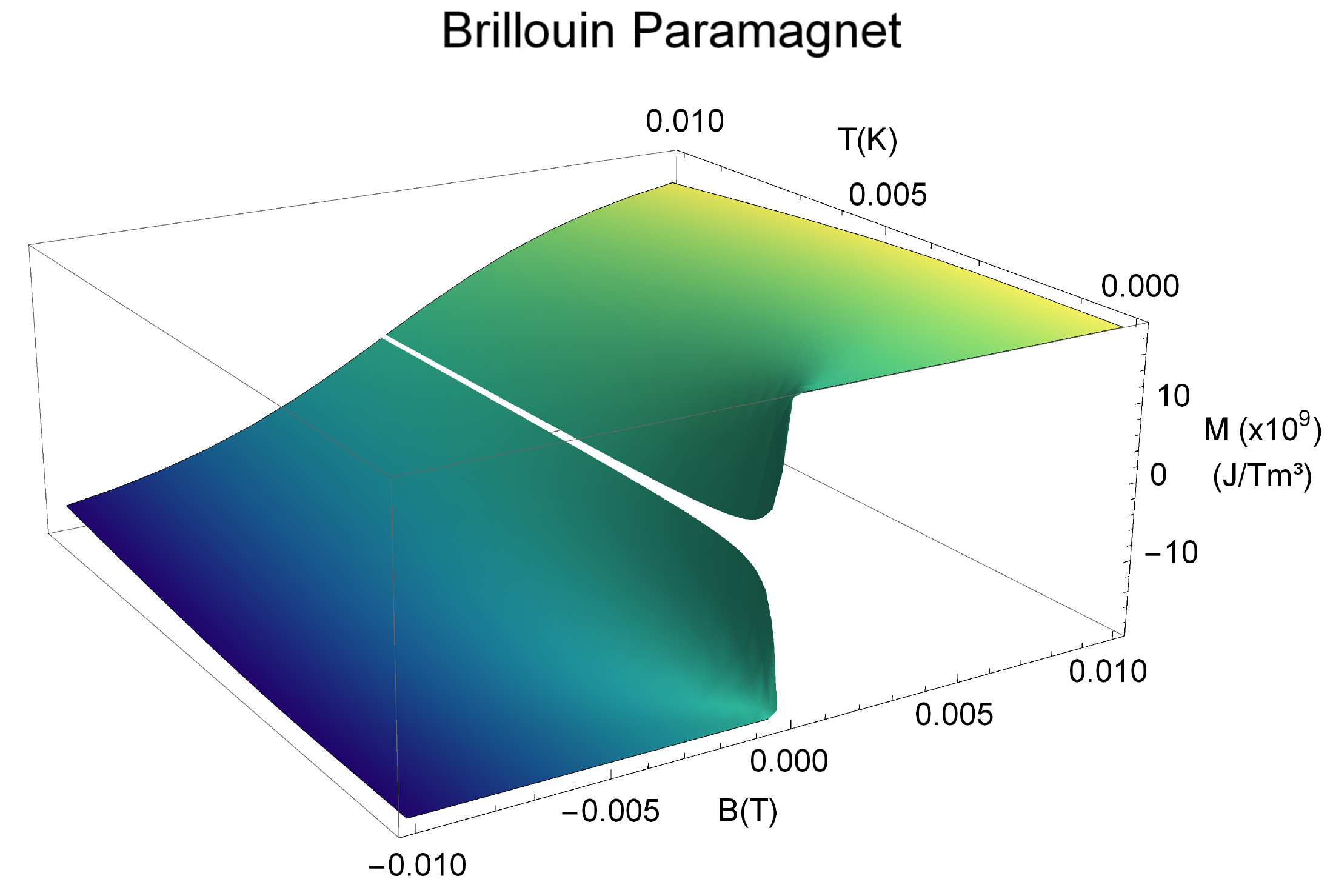}
\caption{\footnotesize Magnetization $M$ as a function of temperature (0 $<$ $T$ $<$  0.01\,K) and magnetic field ($-$0.01\,T $< B < 0.01$\,T) for the classical Brillouin paramagnet.}
\label{Fig-3}
\end{figure}
For arbitrary values of $J$ it can be calculated using the expression \cite{zhu}:
\begin{equation}
\Gamma_{\textmd{mag}} = -\frac{1}{T}\frac{(\partial S/\partial B)_{T}}{(\partial S/ \partial T)_{B}}.
\label{MCE}
\end{equation}
We can calculate the entropy $S$ employing the Helmholtz free energy $F$ per spin:
\begin{equation}
F = -k_{B}T\textmd{ln}[{Z_J(y)}],
\end{equation}
where $Z_J(y)$ is the partition function, given by:
\begin{equation}
Z_J(y) = \frac{\sinh[(2J+1)\frac{y}{2J}]}{\sinh[\frac{y}{2J}]}.
\end{equation}
Thus, the Helmholtz free energy is:
\begin{equation}
F = -k_B T\textmd{ln}\left(\frac{\sinh\left[\frac{(2J+1)y}{2J}\right]}{\sinh\left[\frac{y}{2J}\right]}\right).
\label{helmholtz}
\end{equation}
From Eq.\,\ref{helmholtz}, the entropy $S$ can be easily calculated:
\begin{equation}
S = -\left(\frac{\partial F}{\partial T}\right)_B.
\end{equation}
The resulting expression for the entropy reads:
\begin{equation}
S(y) = k_B\cdot[\textmd{ln}Z_J(y) - yB_J(y)].
\label{entropia}
\end{equation}
Regarding the entropy ($S$) (Eq.\,\ref{entropia}), it is worth recalling the adiabatic demagnetization using a paramagnetic system. Upon applying a magnetic field, the spins are aligned in the direction of $B$ reducing thus the entropy of the system. Then, the magnetic field is removed adiabatically and the temperature of the system decreases. Such a well-known adiabatic demagnetization procedure is frequently employed in order to achieve low-temperatures in the $\mu$K range.

The calculation of the MCE is straightforward and the resulting expression for any $J$ regarding the Brillouin-paramagnet reads:
\begin{equation}
\Gamma_{\textmd{mag}} = \frac{1}{B}.
\label{final}
\end{equation}
From Eq.\,\ref{final} one can again directly conclude that $\Gamma_{\textmd{mag}}$ for the Brillouin-paramagnet depends only on the magnetic field and it diverges as $B \rightarrow 0$ at any temperature.

The effect of the SO interaction on $\Gamma_{\textmd{mag}}$ is discussed in the next section.
\section{The Spin-Orbit Interaction}\label{SO-section}
The interaction between the orbital angular momentum of the nucleus and the electron spin angular momentum is the well-known SO interaction \cite{nolting}. The latter leads to a splitting of the electrons energy levels in an atom. Since the energy levels are affected by the SO interaction, it is of our interest to study the influence of the SO interaction on the Gr\"uneisen parameter.
Thus, in order to take into account the SO interaction, it is necessary to make use of the Hamiltonian, which considers such interaction. For a $S=3/2$ system, such a Hamiltonian was already reported in Refs.\,\cite{Koidl,Sati,Ney} and it has the form:
\begin{equation}
H_{spin} = \mu_B g_{pa} B_z S_z + \mu_B g_{pe} (B_x S_x + B_y S_y) + DS_z^2,
\end{equation}
where $g_{pa}$, $g_{pe}$ and $D$ stand for the gyromagnetic factors of the anisotropic system and the zero-field splitting constant [$D$ = (5.479$\times$10$^{-23}$)\,J = 3.97\,K], respectively. The matrices $S_x$, $S_y$, and $S_z$ can be easily found by the usual operation rules of quantum mechanics. Thus, it is only necessary to diagonalize the resulting operator $H_{spin}$ in order to obtain the eigen-energies. Considering that $B_y=B_z=0$ and $B_x \neq 0$ as reported in Ref.\,\cite{Ney}, one obtains:
$$ H_{spin} = $$
$$ \left(
\begin{array}{cccc}
\frac{9 \hslash ^2 D }{4} & \frac{\sqrt{3} \hslash \mu_B g_{pe} B_x}{2} & 0 & 0 \\ \\
\frac{\sqrt{3} \hslash \mu_B g_{pe} B_x}{2} & \frac{\hslash ^2 D }{4} & \hslash \mu_B g_{pe} B_x & 0 \\ \\
0 & \hslash \mu_B g_{pe} B_x & \frac{\hslash ^2 D}{4} & \frac{\sqrt{3} \hslash \mu_B g_{pe} B_x}{2} \\ \\
0 & 0 &  \frac{\sqrt{3} \hslash \mu_B g_{pe} B_x}{2}  & \frac{9 \hslash ^2 D}{4} \\
\end{array}
\right), $$
and the diagonalization yields four values of eigen-energies given by:
$$ E_1 = \dfrac{1}{2} \mu_B g_{pe} B_x + \dfrac{5}{4} D + \sqrt{\mu_B ^2 g_{pe}^2 B_x ^2 - D g_{pe} \mu_B B_x + D^2}, $$
$$ E_2 = \dfrac{1}{2} \mu_B g_{pe} B_x + \dfrac{5}{4} D - \sqrt{\mu_B ^2 g_{pe}^2 B_x ^2 - D g_{pe} \mu_B B_x + D^2}, $$
$$ E_3 = -\dfrac{1}{2}\mu_B g_{pe} B_x + \dfrac{5}{4} D + \sqrt{\mu_B ^2 g_{pe}^2 B_x ^2 + D g_{pe} \mu_B B_x + D^2}, $$
\begin{equation}
E_4 = -\dfrac{1}{2}\mu_B g_{pe} B_x + \dfrac{5}{4} D - \sqrt{\mu_B ^2 g_{pe}^2 B_x ^2 + D g_{pe} \mu_B B_x + D^2}.
\end{equation}
\begin{figure}[!h]
{%
  \includegraphics[clip,width=\columnwidth]{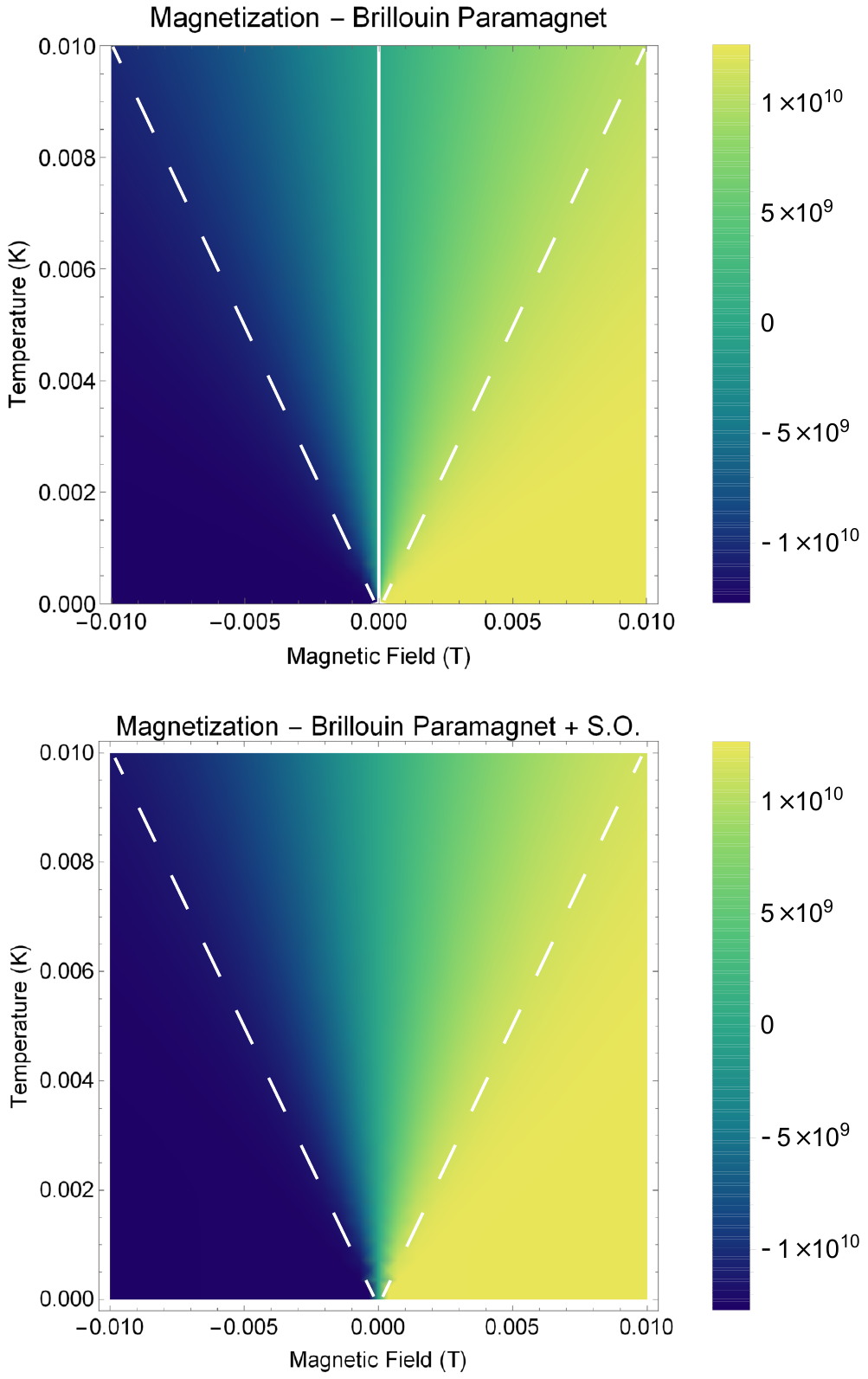}%
}
\caption{\footnotesize Density plots of the magnetization as a function of magnetic field ($-$0.01\,T $< B <$ 0.01\,T) and temperature ($T$ $<$ 0.01\,K) for a Brillouin paramagnet with and without considering the SO interaction. The white dashed lines are a guide to the eyes in order to compare the magnetization cones in both cases.}
\label{Fig-4}
\end{figure}
Yet, the free-energy for the Brillouin-like paramagnet considering the SO interaction reads:
\begin{eqnarray}
\nonumber F &=& -k_B  T \textmd{ln}\left\{2 e^{-\frac{2 B g_{\tiny J} \mu_B +5D}{4 k_B T}} \times\right. \\
\label{freeenergySO} && \times\left[\cosh \left(\frac{\sqrt{B^2 {g_{J}}^2 {\mu_B}^2-B D g_J \mu_B +D^2}}{k_B T}\right)+\right. \\
\nonumber && \left.\left.+\,\,e^{\frac{B g_J \mu_B }{k_B T}} \cosh \left(\frac{\sqrt{B^2 {g_J}^2 {\mu_B}^2+B D g_J \mu_B+D^2}}{k_B T}\right)\right]\right\}.
\end{eqnarray}
Replacing $D$ = 0 in Eq.\,\ref{freeenergySO} and simplifying the resultant expression, it is possible to obtain:
\begin{equation*}
F = - k_B T \textmd{ln}\left[4 \cosh \left(\frac{y}{2J}\right) \cosh \left(\frac{y}{J}\right)\right].
\end{equation*}
Employing the hyperbolic trigonometric identities and again simplifying the equation:
\begin{equation}
F = -k_B T\textmd{ln}\left(\frac{\sinh\left[\frac{2y}{J}\right]}{\sinh\left[\frac{y}{2J}\right]}\right),
\end{equation}
\noindent which is the very same free energy of the Brillouin paramagnet in Eq.\,\ref{helmholtz} employing $J$ = 3/2 without considering the SO interaction. In other words, the classical Brillouin paramagnet is recovered when the zero-field splitting $D$ $\rightarrow$ 0.
\begin{figure}[!h]
\centering
\includegraphics[angle=0,width=\columnwidth]{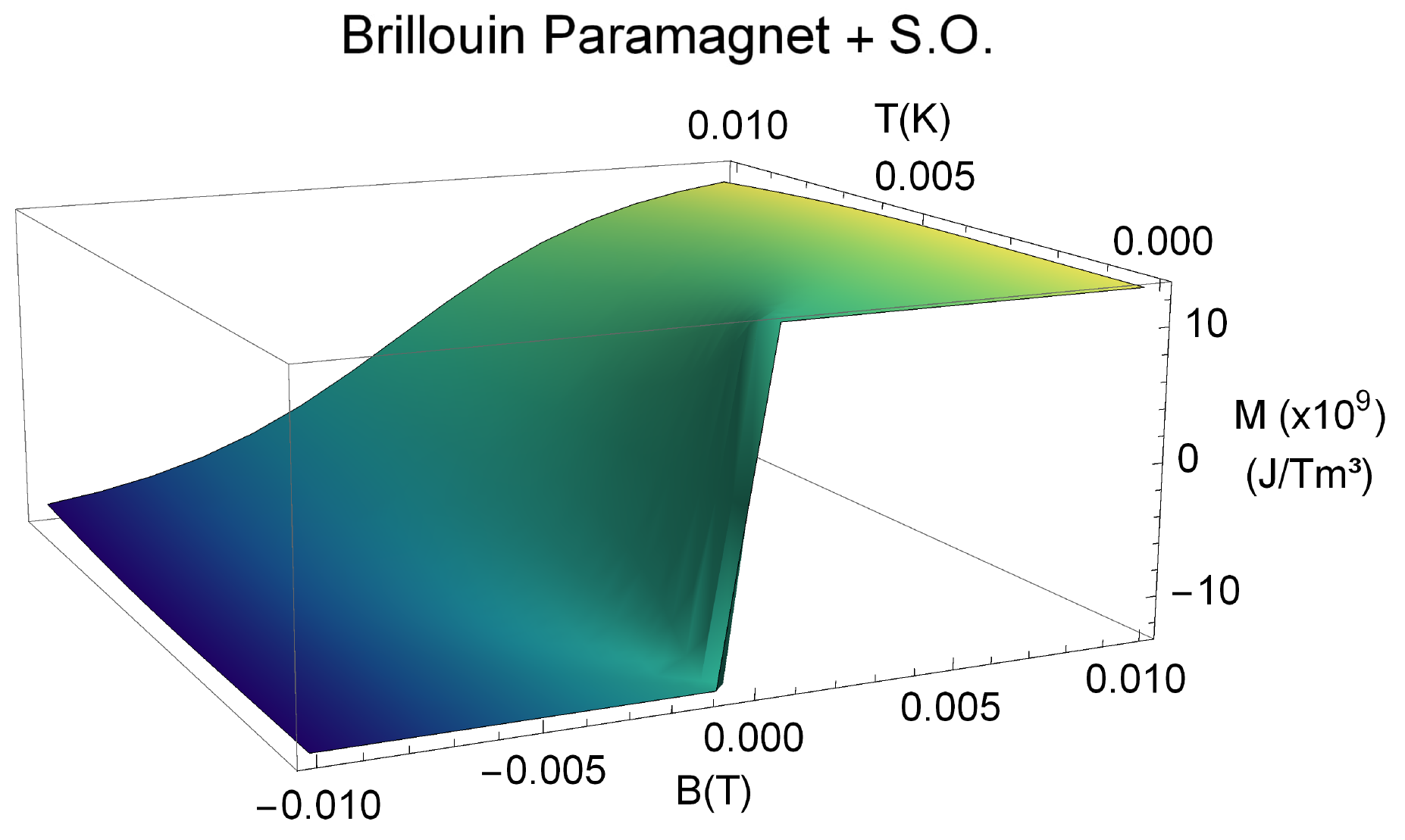}
\caption{\footnotesize Magnetization $M$ as a function of temperature (0 $<$ $T$ $<$  0.01\,K) and magnetic field ($-$0.01\,T $< B < $ 0.01\,T) for the Brillouin paramagnet considering SO interaction. Details in the main text.}
\label{Fig-5}
\end{figure}
\begin{figure}[!htb]
{%
  \includegraphics[clip,width=\columnwidth]{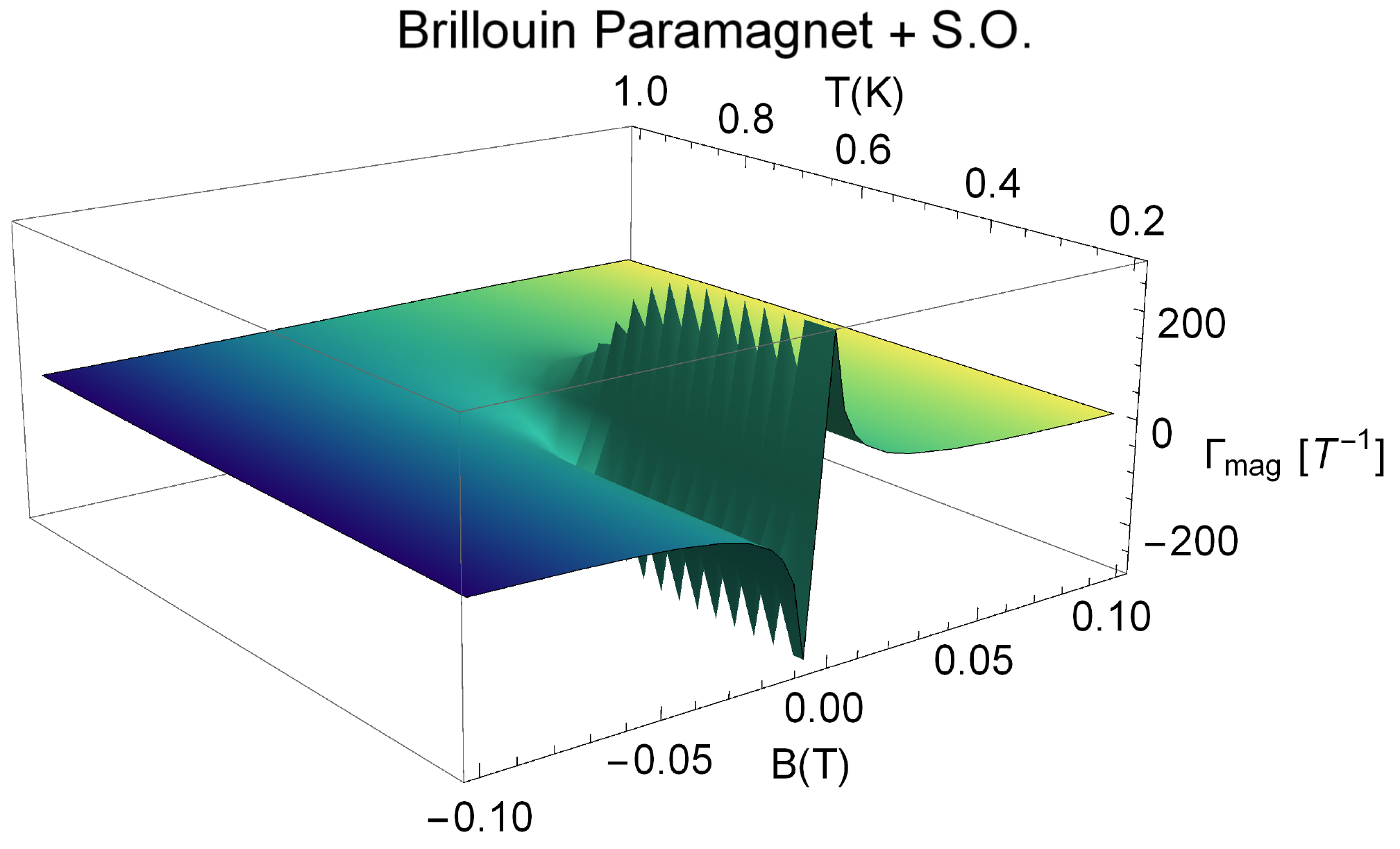}%
}
\caption{\footnotesize 3D-plot of the magnetic Gr\"uneisen parameter $\Gamma_{mag}$ as a function of magnetic field and temperature for the Brillouin paramagnet considering the SO interaction. The depicted \emph{zig-zag} is a consequence of the numerical calculation and thus not intrinsic.}
\label{Fig-6}
\end{figure}
Since the eigen-energies were found, it is then possible to obtain the partition function and consequently the observable quantities, specially the magnetic Gr\"uneisen parameter  $\Gamma_{\textmd{mag}}$ (see Appendix).
In this context, we have performed numerical calculations and made the density and 3D plots of both the magnetization and the magnetic Gr\"uneisen parameter for the Brillouin paramagnet as well as considering the SO interaction.
As can be seen from Fig.\,\ref{Fig-4}, a comparison between the classical Brillouin system for $S$ = 3/2 and SO coupling shows that the magnetization density plot is slightly altered for non-zero $D$. Figure\,\ref{Fig-4} shows that the magnetization is much more sensitive to magnetic field changes for any value of temperature in the case of SO interaction. For the case where the SO interaction is lacking, the magnetization presents a weaker dependence regarding magnetic field changes. From the eigen-energies, we can see that as the magnetic field approaches zero, the degree of degeneracy of the eigenenergies is 2, whereas for the case where no SO interaction is considered (analogously, for $D$ = 0), we have a degree of degeneracy 4 (all the eigenenergies have the same value and equal zero). The magnetic Gr\"uneisen parameter presents a singular behavior in the vicinity of $T$ = 0\,K and $B$ = 0\,T (Fig.\,\ref{Fig-6}).
In other words, $\Gamma_{\textmd{mag}}$ diverges as $B$ $\rightarrow$ 0\,T, a fingerprint of a quantum phase transition. At this point it is important to recall the results reported in Ref.\cite{zhu} obtained using scaling arguments for any QCP tuned by magnetic field:
\begin{equation}
\Gamma_{B,cr}(T \rightarrow 0) = - G \frac{1}{(B-B_c)}, 
\label{Gamma-critico}
\end{equation}
where $cr$ refers to the critical contribution of $\Gamma_B$, $B_c$ is the critical magnetic field and $G$ is an universal pre-factor. Note that Eqs.\,\ref{Gamma-critico} and \ref{final} are quite similar.
The presence of an additional energy scale, namely $D$, gives rise to a temperature-dependent magnetic Gr\"uneisen parameter (cf.\,Fig.\,\ref{Fig-6}) and, it diverges upon approaching $B$ = 0\,T and $T$ = 0\,K, as expected for a quantum critical-like behavior.
\section{The One-Dimensional Ising Model Under Longitudinal Field}\label{Ising-section}
For the 1DI model, all the physical quantities discussed in this section are given per mole of particles. 
For the sake of completeness, we start recalling the 1DI model and its key equations \cite{nolting}, where the Hamiltonian for a linear chain of $N$ spins is expressed by the form:
\begin{equation}
H = \sum_{i=1}^{N}J_{i,i+1}S_i S_{i+1} - B\sum_{i=1}^{N}S_i,
\end{equation}
where $J_{i,i+1}$ is the coupling constant between the $i$- and $i$+1-site spin, and $S_i$ is the spin on the $i$-site. Also, the magnetization is given by:
\begin{equation}
  \label{Mag1D}
  M_{\textmd{1DI}}(T,B) = \mu_B \frac{\sinh\left(\beta \mu_B B\right)}{[\cosh^2\left(\beta \mu_B B\right)-2e^{2\beta J}\sinh(2 \beta J)]^{1/2}},
\end{equation}
where $\beta$ = $1/k_BT$ and $J$ is the coupling constant between two neighbour spins. From Eq.\,\ref{Mag1D} one can deduce that for the 1DI model spontaneous magnetization is not possible, namely $M$($T \neq $ 0, $B$ = 0) = 0. It is now worth analyzing the 1DI magnetic susceptibility, which reads \cite{nolting}:\vspace{0.2cm}
\begin{equation}
  \label{Chi1D}
   \chi_{1DI}(T,B) = \beta \mu_B^2 \frac{\cosh(\beta \mu_B B)(1-2e^{-2 \beta J}{\sinh(2 \beta J)})}{[\cosh^2(\beta \mu_B B)-2e^{-2\beta J}\sinh(2 \beta J)]^{3/2}}.
\end{equation}

\begin{figure}[h]
\centering
\includegraphics[width=\columnwidth]{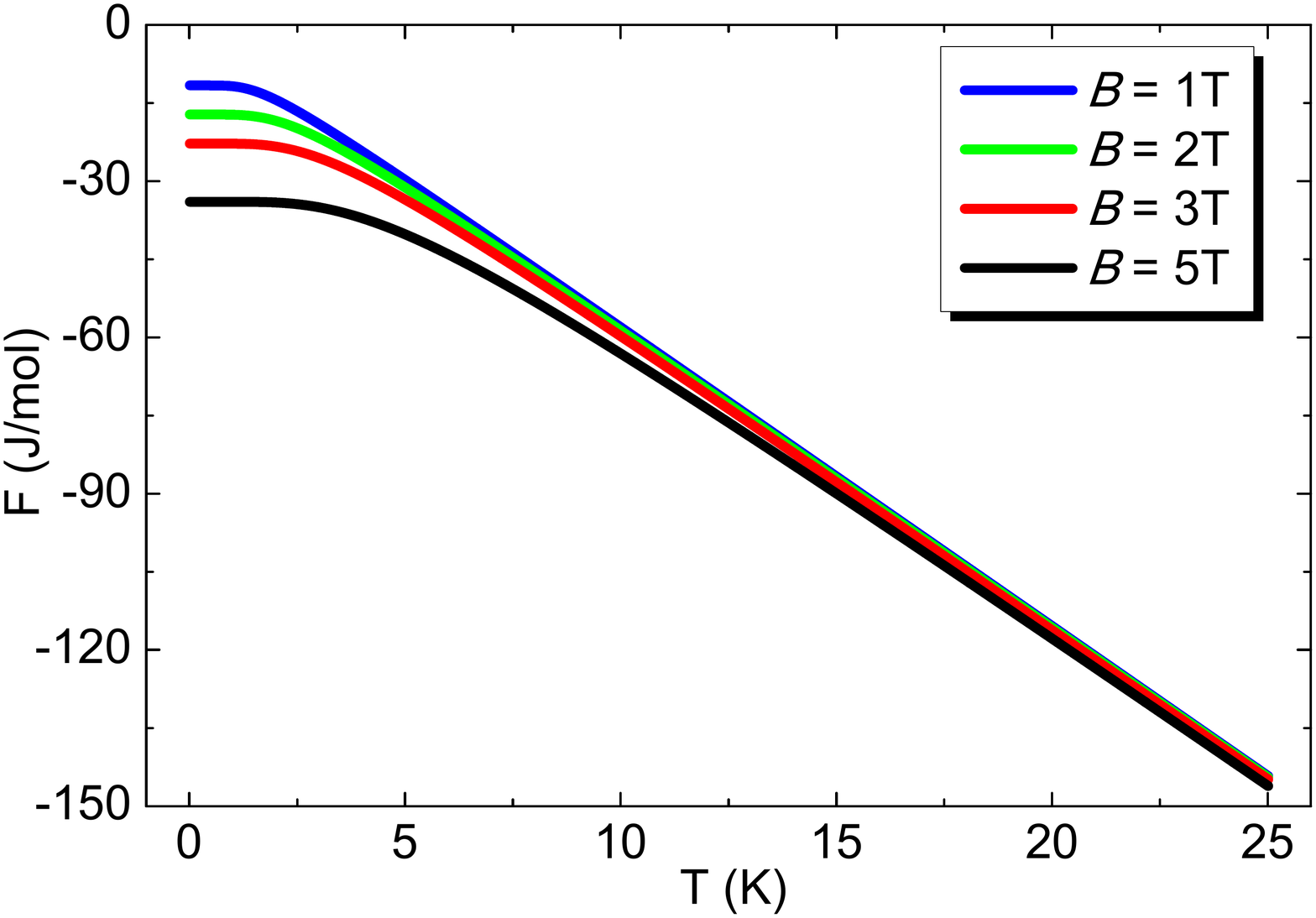}
\caption{\footnotesize The Helmholtz free energy $F$ (Eq.\,\ref{helmholtzfreeenergy}) as a function of temperature for various magnetic fields, cf.\,label. From the data, it can be seen that, as the temperature increases, the free energy decreases. Further details in the main text.}
\label{Fig-7}
\end{figure}

\begin{figure}[h]
\centering
\includegraphics[width=\columnwidth]{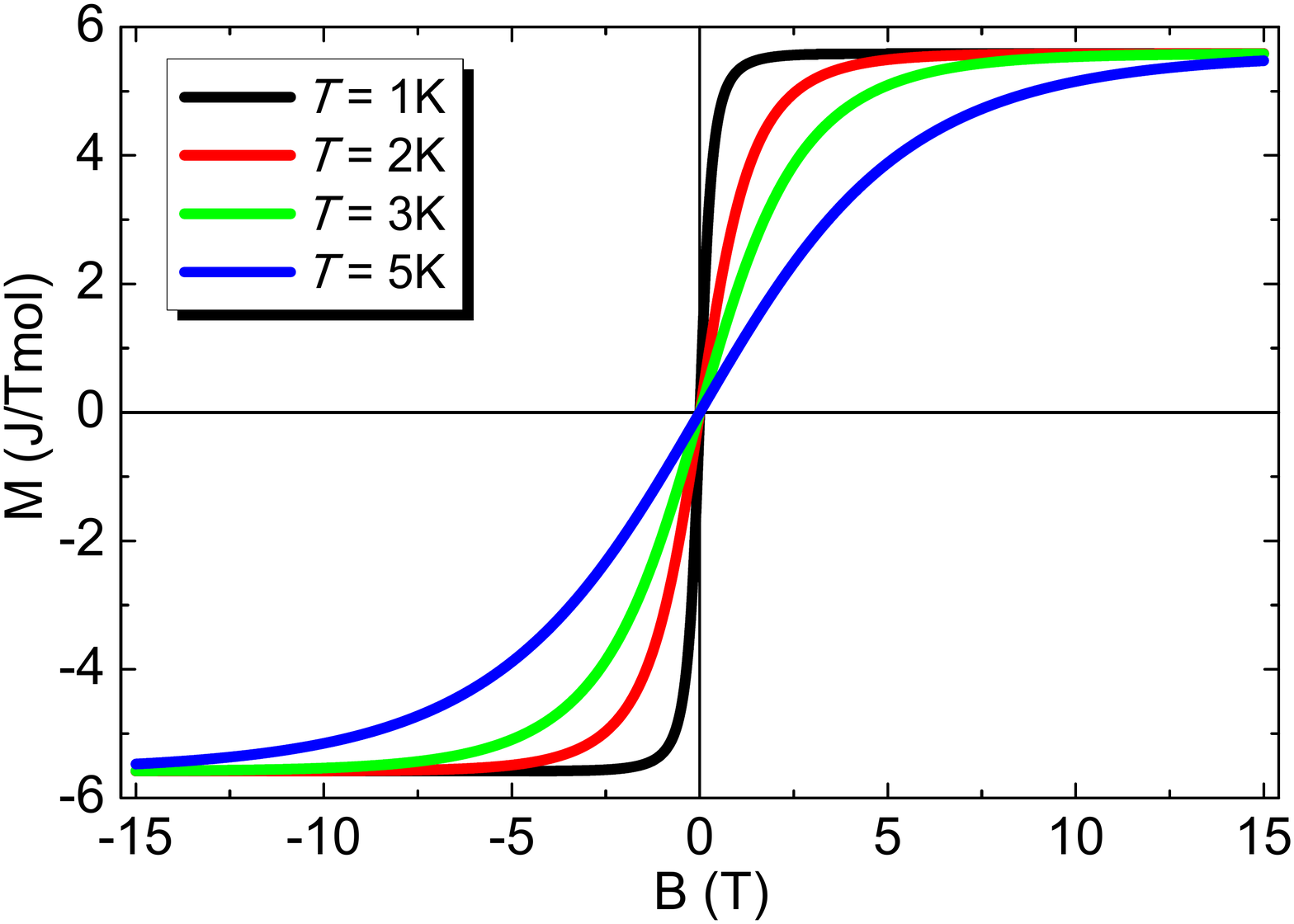}
\caption{\footnotesize Magnetization $M$ as a function of magnetic field $B$ at various temperatures as indicated in the label. Further details are given in the main text.}
\label{Fig-8}
\end{figure}

\begin{figure}[h]
\centering
\includegraphics[width=\columnwidth]{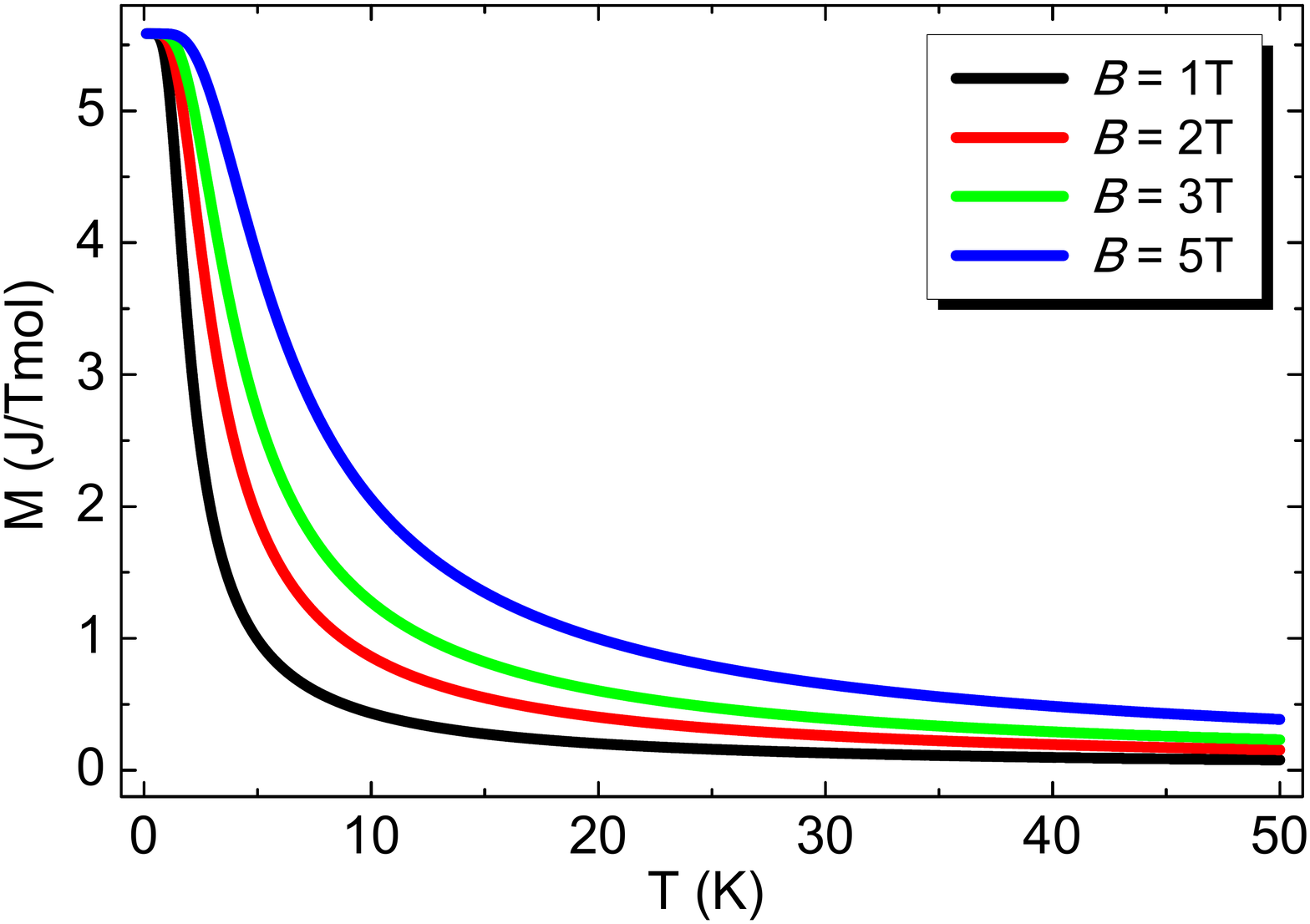}
\caption{\footnotesize Magnetization $M$ as a function of temperature $T$ under various values of magnetic field as indicated in the label. Further details are given in the main text.}
\label{Fig-9}
\end{figure}

For vanishing magnetic field $\chi$($T$, $B \rightarrow$ 0) = $\beta \mu_B^2 e^{2 \beta J}$, \emph{i.e.}, for $B \rightarrow$ 0 and $T \rightarrow$ 0, $\chi$ diverges as expected for a QCP. In other words, for the 1DI model at $T$ = 0\,K a vanishing small external magnetic field suffices to produce long-range magnetic ordering. The specific heat at zero-field is given by:
\begin{equation}
  \label{eq:SH1D}
  C_{B = 0\,\textmd{T}} = k_B \frac{\beta^2 J^2}{\cosh^2(\beta J)}.
\end{equation}
In order to perform an analysis of the 1DI model for generic $B$ and $T$, the Helmholtz free energy is calculated employing the expression \cite{huang}:
\begin{equation}
F(B,T)=-J-k_B T\textmd{ln}\left[\eta+\sqrt{\tau^2+\vartheta}\right],
\label{helmholtzfreeenergy}
\end{equation}
where $\eta$, $\tau$ and $\vartheta$ stand for:
\begin{equation}
\vartheta = \exp \left( \frac{-4J}{k_B T} \right);
\eta=\cosh\left(\frac{\mu_B B}{k_B T}\right);
\tau=\sinh\left(\frac{\mu_B B}{k_B T}\right).
\end{equation}
Figure\,\ref{Fig-7} shows the behavior of the free energy for different applied magnetic fields as a function of temperature. It can be seen that if both $J$ and $B$ are held constant, an increase in the temperature causes a decrease in the free energy.
We also introduce an equivalent definition of the MCE via Maxwell-relations, namely the magnetic Gr\"uneisen parameter \cite{MGarst}:
\begin{equation}
\Gamma_{\textmd{mag}}=-\frac{(\partial M / \partial T )_B} {C_B},
\end{equation}
where:
\begin{equation}
C_B=T \left( \frac{\partial S} { \partial T } \right)_B.
\label{specificheat_eq}
\end{equation}
Since the magnetization was already presented, the calculation of $\Gamma_{\textmd{mag}}$ and the obtainment of the MCE for the 1DI model under longitudinal field is also straightforward.
From Eq.\,\ref{Mag1D} we can see that as $B \rightarrow 0$, also $M \rightarrow 0$, which means that there is no spontaneous magnetization at finite temperature for the 1DI model, as previously stated. Figures \ref{Fig-8} and \ref{Fig-9} show the behavior of the magnetization when $T$ is held constant and $B$ varies and when $B$ is held constant and $T$ varies, respectively.
\begin{figure}[!h]
\centering
\includegraphics[width=\columnwidth]{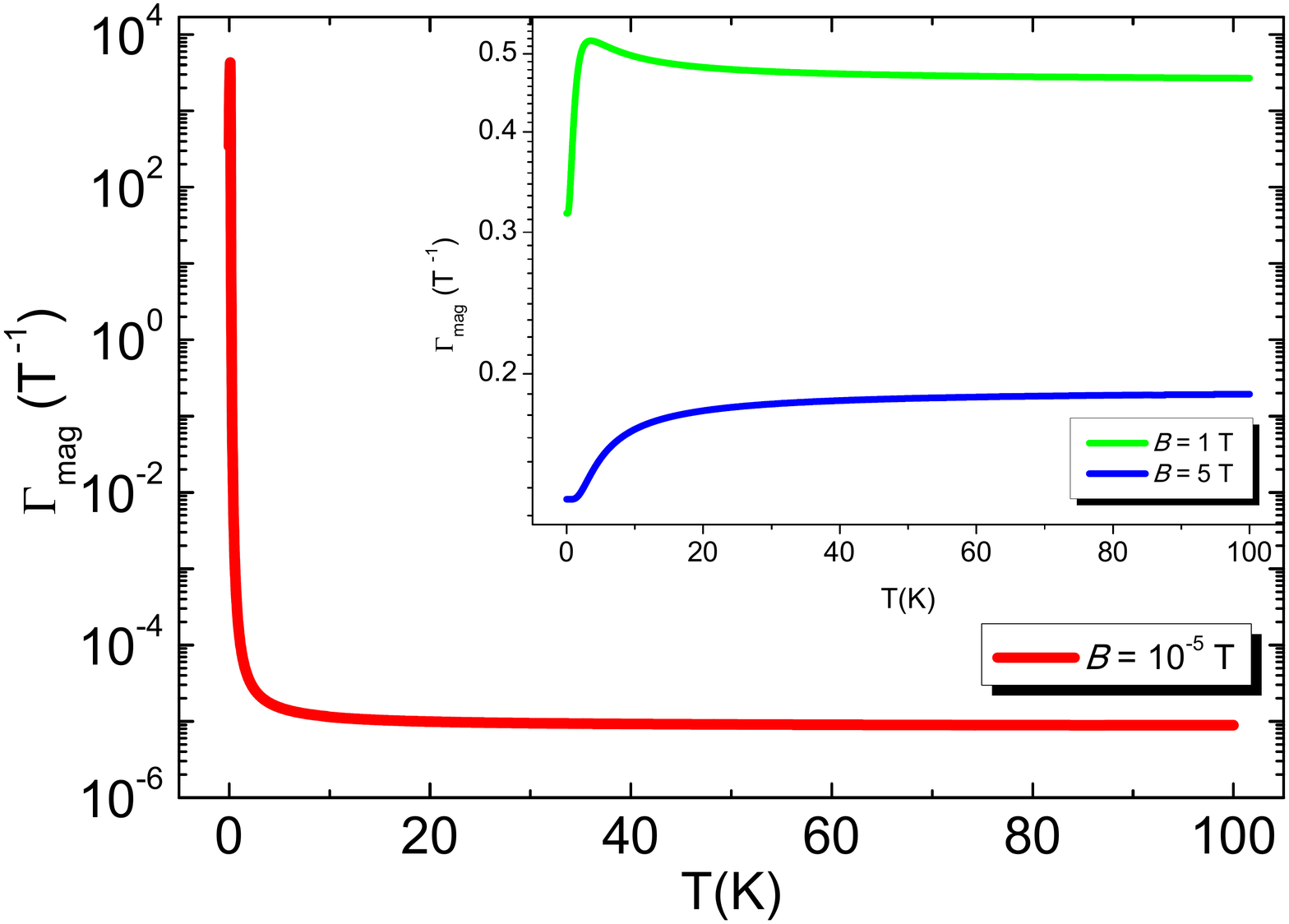}
\caption{\footnotesize Main panel: the magnetic Gr$\ddot{\textmd{u}}$neisen parameter $\Gamma_{\textmd{mag}}$ \emph{versus} $T$ under low-field ($B = $ 10$^{-5}$\,T). Note the logarithmic scale. As $T$ $\rightarrow$ 0\,K $\Gamma_{\textmd{mag}}$ diverges. The absence of a classical phase transition at finite temperatures can be interpreted as a direct consequence of the Mermin-Wagner theorem. Inset: $\Gamma_{\textmd{mag}}$ \emph{versus} $T$ for $B = $ 1 and 5\,T. Further details in the main text.}
\label{Fig-10}
\end{figure}
\begin{figure}[!h]
\includegraphics[width=\columnwidth]{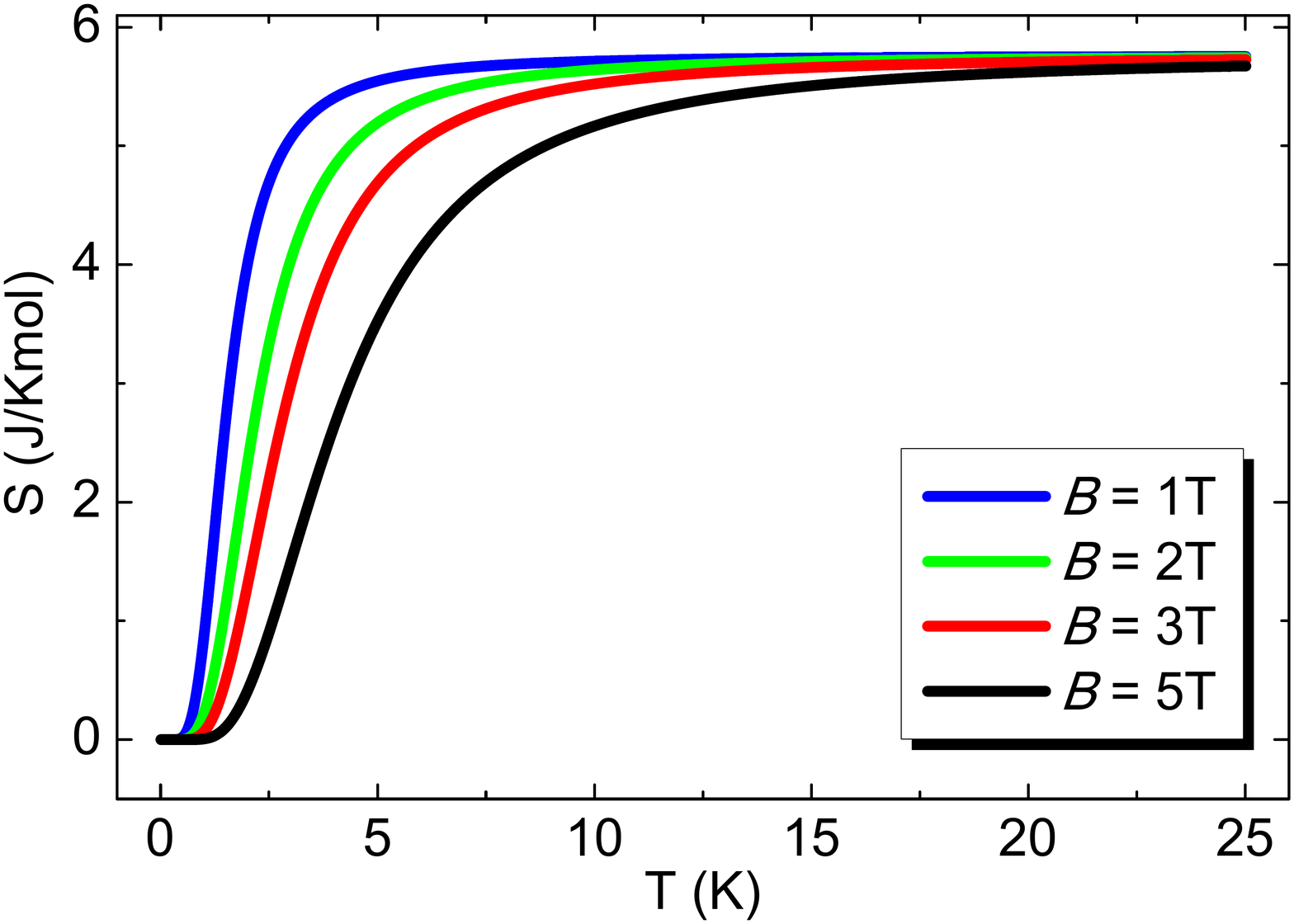}
\caption{\footnotesize Entropy $S$ as a function of temperature $T$ for various values of magnetic field $B$, cf.\,indicated in label. It can be seen that the entropy increases with temperature, which is the expected behavior of the system, \emph{i.e.}, the magnetic disorder is increased upon increasing the thermal energy. Another interesting aspect of the entropy function is its saturation point. When the temperature is sufficiently increased, the entropy reaches a constant value. The required temperature to the saturation increases as the magnetic coupling constant $J$ is increased.}
\label{Fig-11}
\end{figure}
\begin{figure}[!h]
\centering
\includegraphics[width=\columnwidth]{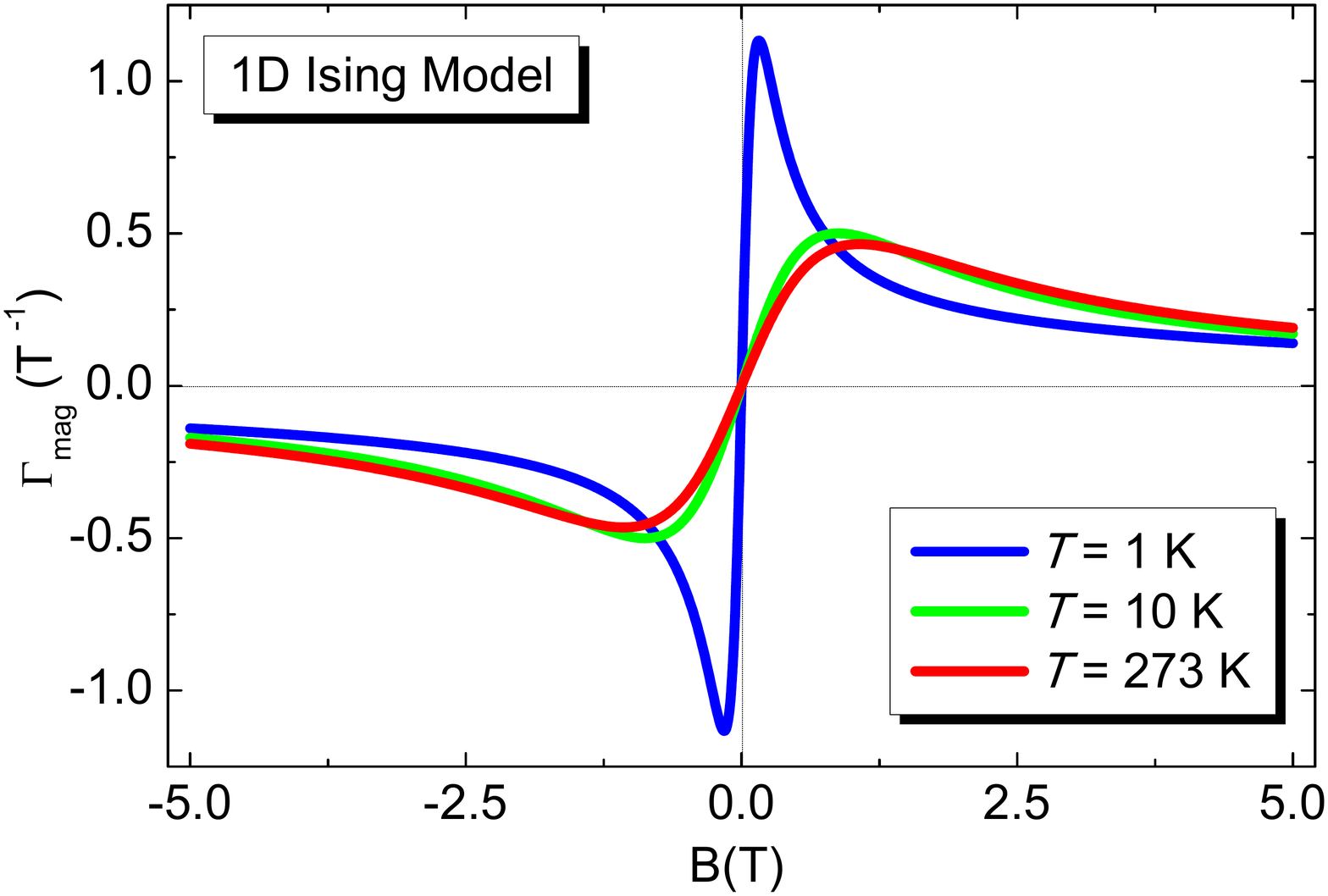}
\caption{\footnotesize The magnetic Gr\"uneisen parameter $\Gamma_{\textmd{mag}}$ as a function of the magnetic field $B$ for various temperatures (see label) for the 1DI model under longitudinal field. By tuning $B$ between positive to negative values, $\Gamma_{\textmd{mag}}$ changes sign.}
\label{Fig-12}
\end{figure}
It is possible already to detect the absence of long-range magnetic order in the system. The Mermin-Wagner theorem ensures that for the 1DI model the spontaneous magnetization is zero for any finite temperature value. If the system would present spontaneous magnetization, in Fig.\,\ref{Fig-8} it would be possible to see a discontinuity in the magnetization at $B=0$ for a certain range of temperature values, given by $T < T_c$, which would be characteristic of a phase transition from ferromagnetic to paramagnetic behavior. However, this behavior is not present in the 1DI Model.
Thus, we can calculate analytically the Gr\"uneisen parameter. The derivatives can be performed straightforwardly yielding:
\begin{equation}
S = k_B \cdot \left[ \frac{ \Sigma}{k_B T \left(\sqrt{\vartheta+\tau^2}+\eta \right) } + \textmd{ln} \left( \sqrt{\vartheta+\tau^2}+\eta \right) \right],
\end{equation}
\begin{equation}
\left( \frac{\partial M} {\partial T} \right)_B = -\frac{2J \tau+B\eta}{T^2\sqrt{\vartheta+\tau^2}\left(k_B \vartheta^{-1}\tau^2+k_B \right)},
\end{equation}
\begin{equation}
C_B=\left[ \frac{ \Sigma}{T \left(\sqrt{\vartheta+\tau^2}+\eta \right) } \right] + \frac{T\Upsilon} {\Phi \varsigma } - \frac{\Theta T\Sigma}  { \varsigma^2},
\end{equation}
where the additional parameters $\Upsilon$, $\Phi$, $\Theta$, $\varsigma$ and $\Sigma$ were introduced purely for compactness, as follows:
$$ \Upsilon = 16 J^2 \tau^2 + 8 J^2 \vartheta - \frac{1}{2} D^2 \vartheta^{-1} \sinh^2\left( \frac{2 \mu_B B}{k_B T} \right) + $$
$$ + D^2\vartheta^{-1}\tau \sinh\left( \frac{2 \mu_B B}{k_B T} \right) \sqrt{\vartheta+\tau^2} + $$
$$ + 2B^2 \cosh \left( \frac{2 \mu_B B}{k_B T} \right) \left( \vartheta^{-1} \tau^2+1 \right) $$
\begin{equation}
+ 2B^2 \eta \sqrt{\vartheta+\tau^2} +4J B \sinh\left( \frac{2 \mu_B B}{k_B T} \right),
\end{equation}
\begin{equation}
\Phi = 2T^2 \sqrt{\vartheta+\tau^2} \left( k_B \vartheta^{-1} \tau^2 + k_B \right),
\end{equation}
\begin{equation}
\Theta = \frac{2\epsilon \vartheta- \frac{1}{2} B \sinh\left( \frac{2 \mu_B B}{k_B T} \right)} {k_B T\sqrt{\vartheta+\tau^2}} + \sqrt{\vartheta+\tau^2} - \frac{B\tau}{k_B T^2},
\end{equation}
\begin{equation}
\varsigma = T \left( \sqrt{\vartheta+\tau^2} + \eta \right),
\end{equation}
\begin{equation}
\Sigma = \frac{\vartheta \left(2J- \frac{1}{2} B\vartheta^{-1} \sinh\left(\frac{2 \mu_B B}{k_B T}\right) \right)} {\sqrt{\vartheta+\tau^2}}-D\tau.
\end{equation}
It can be seen that the MCE for the 1DI model is far from trivial, and we study its behavior by maintaining $B$ and $J$ constant and varying the temperature, cf.\,Fig.\,\ref{Fig-10}. Thus, $\Gamma_{\textmd{mag}}$ for the 1DI model is given by the expression:
\begin{equation}
\Gamma_{\textmd{mag}} =\frac{\frac{2J \tau+B\eta}{T^2\sqrt{\vartheta+\tau^2}\left(k_B \vartheta^{-1}\tau^2+k_B \right)}} {\left[ \frac{ \Sigma}{T \left(\sqrt{\vartheta+\tau^2}+\eta \right) } \right] + \frac{T\Upsilon} {\Phi \varsigma} - \frac{T\Theta \Sigma}  {\varsigma^2}}.
\end{equation}
It is possible to make a Taylor series expansion for $\Gamma_{\textmd{mag}}$ for the case of the 1DI model around $B$ = 0 for a fixed temperature $T$. The obtained expression reads:
\begin{equation}
\Gamma_{\textmd{mag}} (B, T) = \left[ \frac{(e^{2J/k_B T} + 1)^2 \mu_B ^2 (2J + k_B T)}{4J^2 k_B T} \right]B + \mathcal{O}(B^3),
\label{1DIimpar}
\end{equation}
where $\mathcal{O}(B^3)$ represents the higher order terms in the expansion. For $T$ $\rightarrow$ 0\,K, $\Gamma_{\textmd{mag}}$ diverges and a discontinuity takes place at $T$ = 0\,K and $B$ = 0\,T, cf.\,Fig.\,\ref{Fig-12}. From Eq.\,\ref{1DIimpar} it is shown that the mathematical function describing $\Gamma_{\textmd{mag}}$ for the 1DI model is odd with respect to the magnetic field $B$. Hence, Eq.\,\ref{1DIimpar} also explains the obtained symmetrical behavior of $\Gamma_{\textmd{mag}}$ upon varying the magnetic field $B$ from negative to positive values, cf.\,Fig.\ref{Fig-12}.
In order to analytically demonstrate the asymptotic equivalence of $\Gamma_{\textmd{mag}}$ for the 1DI and the Brillouin model, it is possible to make a change in the temperature variable $T$ to 1/$T$ and make an expansion in a Taylor series around 1/$T$ $\rightarrow$ 0 \cite{arfken}. Thus, the expression of $\Gamma_{\textmd{mag}}$ for the 1DI model is simplified and is expressed by:
\begin{equation}
\Gamma_{\textmd{mag}} (B,T \rightarrow \infty) = \frac{\mu _B^2 B}{J^2 + \mu _B ^2 B^2}.
\label{1DIexpanded}
\end{equation}
In this context, when $J$ = 0, the $\Gamma_{\textmd{mag}}$ for the Brillouin model is elegantly recovered, namely 1/$B$ (Eq.\,\ref{final}). This means that at high-temperatures and upon neglecting the magnetic coupling between the nearest neighbors spins, we have shown analytically that the two magnetic models are equivalent. Similarly, it is possible to make the very same expansion for $\Gamma_{\textmd{mag}}$ in the case of the Brillouin model upon considering the SO interaction:
\begin{equation}
\Gamma_{\textmd{mag}} (B,T \rightarrow \infty) = \frac{5{g_J}^2\mu _B^2 B}{4D^2 + 5 {g_J}^2 \mu _B^2 B^2}.
\label{SOexpanded}
\end{equation}
Upon comparing Eqs.\,\ref{1DIexpanded} and \ref{SOexpanded}, it is clear that both expressions have the same mathematical structure except for the constant term $J^2$. Making $J = \sqrt{4/5}D/{g_J}$ and replacing it in Eq.\,\ref{1DIexpanded}, it can be shown that the two models are also equivalent in the asymptotic regime, namely for 1/$T$ $\rightarrow$ 0. Such mathematical similarity in the regime of high-temperatures is a direct consequence of the dominant effects from thermal fluctuations and obviously this does not mean that the models are physically equivalent. Yet, still considering Eq.\,\ref{SOexpanded} it is clear that for $D$ $\rightarrow$ 0, Eq.\,\ref{final} is nicely restored.
Summarizing the results obtained for the longitudinal 1DI model: \textit{i}) it does not present any classical critical behaviour for any finite value of temperature as a consequence of the Mermin-Wagner theorem, and \textit{ii}) for $T \rightarrow 0$\,K, the system shows intrinsic quantum critical behaviour. It is important to emphasize that the Gr\"uneisen parameter cannot be calculated at $T$ exactly equal to $0$\,K, since the corresponding function is not determined in this point.
\section{Concluding Remarks}
Gr\"uneisen in 1908 realized that the volume dependence of the vibrational energy must be taken into account  in order to explain  thermal expansion.  Slowly but steadily the Gr\"uneisen parameter has been incorporated as a thermodynamic coefficient, both in thermodynamical textbooks and in experimental physics; and when measured it can provide information on other  thermodynamic coefficients, as shown in references \cite{EPJ,Stanley}. As a second step from our previous work \cite{EPJ}, we have considered the magnetic analog of the Gr\"uneisen parameter as  a tool  to further probe magnetic systems, in particular at low temperature, when quantum phase transitions are relevant.
In the Introduction we mentioned several exotic manifestations of matter where the Gr\"uneisen parameter could be measured.  Following the Introduction we computed this parameter for several known theoretical models, namely:  the Brillouin paramagnet, yielding a temperature independent Gr\"uneisen parameter, proportional to the inverse of the applied magnetic field; SO interaction model yields a diverging Gr\"uneisen parameter as the temperature goes to zero; and the longitudinal 1D Ising model, known not to exhibit any kind of phase transitions at finite temperature, as a consequence of the well-known Mermin-Wagner theorem. Nevertheless, a quantum phase transition can be inferred. Also we did find an equivalence at high temperatures  between the 1D Ising model with a Brillouin paramagnet with vanishing coupling constant $J$, and with or without a SO interaction included on the latter. Thus, the magnetic Gr\"uneisen parameter can bee seen as a smoking gun  when we probe critical points. Future work will consider other systems, where the thermodynamic coefficients are not readily computed in a complete analytic fashion.
\vspace{0.8cm}
\section{Appendix}
For the Brillouin-like paramagnet considering SO interaction the magnetic Gr\"uneisen parameter $\Gamma_{\textmd{mag}}$ reads:
\begin{equation} \label{eq:5}
    \Gamma_{\textmd{mag}} = \frac{a(T,B)}{c(T,B)},
\end{equation}
where:
\begin{align*}
&a(T,B) = -g_J\mu_B(\sinh[A](F(2D^{2}-3BDg_J\mu_B+4B^{2}{g_J}^{2}\\
&{\mu_B}^{2})\cosh[C]+2Bg_J\mu_B(D^{2}+2B^{2}{g_J}^{2}{\mu_B}^{2})\sinh[C])-G\\
&\cosh[A](2D^{2}\sinh[C]+Bg_J\mu_B(6F\cosh[C]+(3D+4B\\
&g_J\mu_B)\sinh[C]))-2FG(2Bg_J\mu_B\cosh[(Bg_J\mu_B)/(k_B T)] \\
&+D\sinh[(Bg_J\mu_B)/(k_B T)]))
\end{align*}
\begin{align*}
&c(T,B) = 2 F G (\cosh[A] ((2 D^2 + 3 B^2 {g_J}^2 {\mu_B}^2) \cosh[C] + 2 \\
&B F g_J \mu_B \sinh[C]) + 2 ((D^2 + B^2 {g_J}^2 {\mu_B}^2) \cosh[(B g_J \mu_B)/ \\
&(k_B T)] - G \sinh[A] (B g_J \mu_B \cosh[C] + F \sinh[C]) + BD g_J  \\
&\mu_B \sinh[(B g_J \mu_B)/(k_B T)]))
\end{align*}
\begin{equation*}
A = \frac{\sqrt{B^2{g_J}^2{\mu_B}^2 - Bg_J\mu_BD + D^2}}{k_B T}
\end{equation*}

\begin{equation*}
C = \frac{\sqrt{B^2{g_J}^2{\mu_B}^2 + Bg_J\mu_BD + D^2}}{k_B T}
\end{equation*}
\begin{equation*}
F = \sqrt{B^2{g_J}^2{\mu_B}^2 + Bg_J\mu_BD + D^2}
\end{equation*}
\begin{equation*}
G = \sqrt{B^2{g_J}^2{\mu_B}^2 - Bg_J\mu_BD + D^2}
\end{equation*}
\vspace{0.5cm}
\section*{Acknowledgements}
MdS acknowledges financial support from the S\~ao Paulo Research Foundation -- Fapesp (Grants No. 2011/22050-4), National Council of Technological and Scientific Development -- CNPq (Grants No.\,302498/2017-6), the Austrian Academy of Science \"OAW for the JESH fellowship, Prof.\,Serdar Sariciftci for the hospitality. AN and MdS gratefully acknowledge funding by the Austrian Science Fund
(FWF)-Project No. P26164-N20 during the early stage of this work.
\medskip
\bibliographystyle{unsrt}
\bibliography{References-MCE}
\end{document}